\documentclass[aps,prl,reprint,showpacs,groupedaddress]{revtex4-1}
\usepackage{amsmath}
\usepackage{amsfonts}
\usepackage{amssymb}
\usepackage{color}
\usepackage{blindtext}
\usepackage{framed}
\usepackage{graphicx}
\usepackage{bm}
\usepackage{natbib}
\usepackage{latexsym}

\begin{document}

\title{ {\it {\small Published in Physics of Plasmas 23, 100702 (2016)}} 
\\
$\,$
\\
General Theory of the Plasmoid Instability}

\author{L. Comisso}
\email{lcomisso@princeton.edu}
\author{M. Lingam}
\author{Y.-M. Huang}
\author{A. Bhattacharjee}

\affiliation{Department of Astrophysical Sciences and Princeton Plasma Physics Laboratory, Princeton University, Princeton, NJ 08544, USA}

\begin{abstract}
A general theory of the onset and development of the plasmoid instability is formulated by means of a principle of least time. The scaling relations for the final aspect ratio, transition time to rapid onset, growth rate, and number of plasmoids are derived and shown to depend on the initial perturbation amplitude $\left({\hat w}_0\right)$, the characteristic rate of current sheet evolution $\left(1/\tau\right)$, and the Lundquist number $\left(S\right)$. They are \emph{not} simple power laws, and are proportional to $S^{\alpha} \tau^{\beta} \left[\ln f(S,\tau,{\hat w}_0)\right]^\sigma$. The detailed dynamics of the instability is also elucidated, and shown to comprise of a period of quiescence followed by sudden growth over a short time scale. 
\end{abstract}

\maketitle

The rapid conversion of magnetic energy into plasma particle energy through the process of magnetic reconnection is of great importance in the realm of plasma physics and astrophysics \cite{TajimaShibata1997,Biskamp_2000,Kulsrud_2005,Zweibel_2009}. Sawtooth crashes, magnetospheric substorms, stellar and gamma-ray flares are just a few examples of pheneomena in which magnetic reconnection plays an essential role.

In large systems, such as those found in space and astrophysical environments, the potential formation of highly elongated current sheets would result in extremely low reconnection rates, which fail to account for the observed fast energy release rates \cite{Retino_2007,Slavin_2009,Su_2013}. However, such current sheets are subject to a violent linear instability that leads to their breakup, giving rise to a tremendous increase in the reconnection rate that appears to be very weakly dependent on the Lundquist number of the system in the nonlinear regime \cite{BHYR_2009,Daugh_2009,Cassak_2009,Huang_2010,Huang_2011,Lou_2012,Murphy_2013,Huang_2013,Comisso2015,Ni2015}. This crucial instability, which serves as a trigger of fast reconnection, is the plasmoid instability \cite{Biskamp_2000}, thus dubbed as it leads to the formation of plasmoids.

In the widely studied Sweet-Parker current sheets, which are characterized by an inverse aspect ratio $a/L \sim S^{-1/2}$, Tajima and Shibata \cite{TajimaShibata1997}, as well as Loureiro \emph{et al.} \cite{Lou_2007}, have found that the growth rate $\gamma$ and the wavenumber $k$ of the plasmoid instability obey $\gamma {\tau_A} \sim S^{1/4}$ and $kL \sim S^{3/8}$, where $\tau_A$ is the Alfv{\'e}nic timescale based on the length of the current sheet. Since the Lundquist number $S$ is extremely large in most space and astrophysical plasmas \cite{JiDau2011}, the linear growth of the instability turns out to be surprisingly fast, and the number of plasmoids produced is also very high. Other notable works have since followed, which have verified and extended the work on the plasmoid instability in different contexts \cite{SLUSC_2009,Ni_2010,Baalrud_2012,LSU_2013,Comisso2016}.

Despite the success of the theory, its limitations soon became evident. For sufficiently high growth rates, Sweet-Parker current sheets cannot be attained as current layers are linearly unstable and disrupt before this state is achieved. In order to bypass this limitation, Pucci and Velli \cite{Pucci_2014} conjectured that current sheets break up when $\gamma {\tau_A} \sim 1$. Later, Uzdensky and Loureiro \cite{UzdLou2016} considered a similar criterion ($\gamma \tau = 1$) as the end-point of the linear stage of the instability, presenting an appealing but heuristic discussion for the case of a current sheet evolving on the timescale $\tau$. However, as demonstrated below, the most interesting physics occurs in the regime $\gamma \tau > 1$. Yet, a quantitative theory that encompasses all of the intricate physics underlying the onset and development of the plasmoid instability has been elusive. It is the purpose of this Letter to develop a quantitative theory of the plasmoid instability in time evolving current sheets based on a principle of least time. We obtain new and surprising results, such as scalings that are not power laws, and provide a detailed picture of how the plasmoid instability arises and develops.



In a time evolving current sheet, tearing modes become unstable at different times and exhibit different instantaneous growth rates ${\gamma}(k,t)$. Their amplitude changes in time according to ${\psi}(k,t) = {\psi_{0}} \exp \big (\int_{t_0}^t {{\gamma}(t')dt'} \big)$, where $\psi$ represents the tearing eigenfunction and ${\psi_{0}} := {{\psi}{(k,t_0)}}$. During the linear evolution, the amplitudes of the different modes are small, thereby ensuring that they don't affect each other and the current sheet evolution. Their linear evolution ends when the plasmoid half-width \cite{Biskamp_2000}
\begin{equation} \label{half_width}
{w}(k,t) = 2{\left( {{\psi}a/{B_0}} \right)^{1/2}}
\end{equation}
grows to the same order of the inner resistive layer width 
\begin{equation} 
{\delta_{{\rm{in}}}}(k,t)  = {\big[ { \eta {\gamma}{a^2}/{(k{v_A})^2} } \big]^{1/4}} \, .
\end{equation}
Note that ${\psi}$ is now taken to be implicitly evaluated at the resonant surface, ${B_0}$ is the reconnecting magnetic field upstream of the current sheet, and $a$ is the current sheet half-width. Here, $w$ may be regarded as a label of the perturbation amplitude and represents the plasmoid width only if the associated mode is dominant with respect to the others (or, less restrictively, for multiple dominant modes, as long as they are few in number and sufficiently localized in the spectrum). This does turn out to be the case at the end of the linear phase.

In such a complex scenario, we intend to determine the mode (${k_*}$) that emerges from the linear phase and transitions into the nonlinear phase, and the time (${t_*}$) at which this transition occurs. The most general way to address this question arguably relies on the formulation of a principle of least time for the plasmoid instability, i.e., {\it the mode of the plasmoid instability that emerges from the linear phase is the one that traverses it in the least time.} Mathematically, to implement our formulation, we introduce the function
\begin{equation} 
F(k,t):= \delta_{{\rm{in}}}(k,t) - w(k,t) \, .
\end{equation}
It is self-evident that ${w}({k,t_0}) \ll {\delta_{{\rm{in}}}}({k,t_0})$ is necessary for the linear evolution to exist, viz., the initial mode amplitude has to be sufficiently small in large-$S$ plasmas. 
We assume a continuous spectrum of wavenumbers $k$. Then, the least time principle yields the mathematical relations
\begin{equation} \label{Fermat_Principle}
{\left. F(k,t) \right|_{{k_*},{t_*}}} = 0 \, , \quad  {\left. {\frac{{\partial F(k,t)}}{{\partial k}}} \right|_{{k_*},{t_*}}} = 0 \, .
\end{equation}
The second relation is equivalent to the condition $d{t_*}/d{k_*} = 0$, where ${t_*} = {t_*}({k_*})$ implicitly follows from the first relation. This allows us to interpret ${t_*}$ as the variable that is extremized. In addition, it can also be shown \emph{a posteriori} that $\big| (k/\gamma) (\partial \gamma/ \partial k) \big| \ll 1$, and consequently $\delta_{{\rm{in}}} \propto k^{-1/2}$, holds true in the neighborhood of $k_*$. Since $w$ is localized and has a much stronger dependence than $\delta_{{\rm{in}}}$ on $k$, the mode that completes the linear phase in the least time is the dominant (larger amplitude) one that enters the nonlinear phase.

In our subsequent discussion, it is convenient to work with dimensionless quantities. In particular, we normalize all the lengths to the current sheet half-length $L$, the magnetic field to the upstream field ${B_0}$, and the time to the Alfv{\'e}n time $\tau_A = L/v_A$. We use carets to indicate the dimensionless quantities -- $\hat a = a/L$, ${{\hat \delta }_{{\rm{in}}}} = \delta_{{\rm{in}}}/L$, $\hat k = kL$, $\hat \psi  = \psi /L{B_0}$, $\hat t = t/{\tau _A}$, and $\hat \gamma  = \gamma {\tau_A}$. Therefore, the normalized magnetic diffusivity corresponds to the inverse of the Lundquist number, i.e. ${{\hat \eta }^{-1}} = S := {v_A}L/\eta $.

Although our framework is altogether general, we are interested in the case where $L$ and $B_0$ remain approximately constant, while the current sheet width decreases in time via $\hat a(\hat t) = {{\hat a}_0}f(\hat t)$. Here, $f(\hat t)$ is a function that must obey $f({{\hat t}_0}) = 1$ and ${\lim _{\hat t \to \infty }}f(\hat t) = \hat a_0^{-1} S^{-1/2}$. Indeed, $\hat a = S^{-1/2}$ is the natural lower limit to the thickness of a reconnection layer, due to the increase in the Ohmic heating when $B_0 / a$ increases \cite{Kulsrud_2005}.

Now, we can explicitly rewrite the first of Eq. (\ref{Fermat_Principle}) as
\begin{equation} \label{Eq_1a}
{\left. {\left\{ {\ln \left( {\frac{{\hat a_0^{1/2}{{\hat \gamma }^{1/4}}{S^{ - 1/4}}}}{{{{\hat w}_{0}}{{\hat k}^{1/2}}}}} \right) - \frac{1}{2}\int_{{{\hat t}_0}}^{\hat t} {\hat \gamma ({\hat t}')d{\hat t}'} } \right\}} \right|_{{{\hat k}_*},{{\hat t}_*}}} = 0 \, ,
\end{equation}
where ${{\hat w}_{0}}: = {{\hat w}}({\hat k},{\hat t}_0) = 2 {\big( {{\hat \psi }_{0}}{{\hat a}_0} \big)^{1/2}}$ is the label for the initial perturbation amplitude. We assume that the initial perturbation is the same for all wavelengths, but other possibilities can be easily handled with our treatment. Then, we combine the second of Eq. (\ref{Fermat_Principle}) with Eq. (\ref{Eq_1a}) to obtain the {\it least time equation}
\begin{equation} \label{Fderdef}
{\left. {\left\{ {\left( {\hat \gamma \bar t - \frac{1}{2}} \right)\frac{{\partial \hat \gamma }}{{\partial \hat k}} + \frac{{\hat \gamma }}{{\hat k}}} \right\}} \right|_{{{\hat k}_*},{{\hat t}_*}}} = 0 \, ,
\end{equation}
where  $\bar t = \partial /\partial \hat \gamma \int_{{\hat t}_0}^{\hat t} {\hat \gamma (\hat t')d\hat t'}$. This equation leads us to ${\hat k}_*$, $\hat \gamma({{\hat k}_*},{{\hat t}_*}) := {{\hat \gamma}_*}$ and ${{\hat \delta }_{{\rm{in}}}}({{\hat k}_*},{{\hat t}_*}) := {{\hat \delta }_{{\rm{in}}*}}$ as a function of the inverse aspect ratio $\hat a({{\hat t}_*}) := {{\hat a}_*}$.

The procedure delineated above is fairly general, and can be applied to tearing modes with arbitrary degrees of collisionality. However, to proceed further, we must specify an expression for ${\hat \gamma}$. Here, for the sake of definiteness, we focus on resistive tearing modes. When the evolution of the current sheet width is slow, i.e., ${{\hat a}^{ - 1}}d\hat a/d\hat t < {\hat \gamma}$, the growth rate of the tearing mode can be computed using the instantaneous value of ${\hat a}$. For resistive tearing modes, two simple algebraic relations exist, which are valid in two different regimes, depending on the value of the tearing stability parameter ${{\hat \Delta}'}$ \cite{FKR_1963}. In the small-${{\hat \Delta}'}$ regime, defined as ${{\hat \Delta}'} {{\hat \delta }_{{\rm{in}}}} \ll 1$, the classic analysis by Furth, Killeen, and Rosenbluth \cite{FKR_1963} demonstrated that 
\begin{equation} \label{FKRDisp}
\hat{\gamma}_s \simeq c_{\Gamma} \hat{k}^{2/5} \hat{a}^{-2/5} S^{-3/5} {\hat \Delta '^{4/5}} \, ,
\end{equation} 
where $c_{\Gamma}  = {\big[ {{{(2\pi )}^{ - 1}}\Gamma (1/4)/\Gamma (3/4)} \big]^{4/5}} \approx 0.55$. On the other hand, in the large-${{\hat \Delta}'}$ regime, defined as ${{\hat \Delta}'} {{\hat \delta }_{{\rm{in}}}} \gtrsim 1$, Coppi {\it et al.} \cite{Coppi_1976} showed that the growth rate becomes independent of ${{\hat \Delta}'}$, resulting in 
\begin{equation} \label{CoppiDisp}
\hat{\gamma}_l \simeq \hat{k}^{2/3} \hat{a}^{-2/3} S^{-1/3} \, .
\end{equation}
As we are interested in the entire domain of ${{\hat \Delta}'}$, we seek an expression for $\hat{\gamma}$ that (i) is a reasonable approximation of the exact growth rate \cite{Coppi_1976}, (ii) reduces to (\ref{FKRDisp}) and (\ref{CoppiDisp}) in the appropriate limits, and (iii) is simple enough to be analytically tractable. For this purpose we consider the half-harmonic mean of this two relations, i.e., 
\begin{equation} \label{fullgamma}
\hat \gamma  = {{\hat \gamma }_s}\,{{\hat \gamma }_l}/({{\hat \gamma }_s} + {{\hat \gamma }_l}) \, .
\end{equation}
Since the harmonic mean is a Schur-concave function \cite{HLP52}, in addition to being quite simple, we have verified that it fulfills all of the criteria described above. Furthermore, choosing a different approximation for $\hat \gamma$ such as the simpler one employed in \cite{TajimaShibata1997,BHYR_2009,LSU_2013,Huang_2013,Pucci_2014} leads to the same scaling relations presented below, albeit with slightly different numerical factors.

Considering a common Harris-like current sheet, it is known that \cite{FKR_1963} ${{\hat \Delta}'} \hat{a} = 2 \big[(\hat{k}\hat{a})^{-1} - \hat{k}\hat{a}\big]$. As we are not interested in the very slow-growing part of the mode evolution, we consider the regime $\hat k \hat a \ll 1$. Then, without any further simplification, it is possible to obtain the following expression from Eq. (\ref{Fderdef}):
\begin{eqnarray} \label{Fderdf2}
 && \frac{1}{{{{\bar t}_*}}} \Big( 1 + \frac{5}{{14}}{{\hat k}_*}^{ - 16/15}{{\hat a}_*}^{ - 4/3}{S^{ - 4/15}} + \frac{9}{{14}}{{\hat k}_*}^{16/15}{{\hat a}_*}^{4/3}{S^{4/15}} \Big)   \nonumber\\
 &&\qquad= \frac{3}{{14}}{{\hat k}_*}^{2/3}{{\hat a}_*}^{ - 2/3}{S^{ - 1/3}} - \frac{5}{{14}}{{\hat k}_*}^{ - 2/5}{{\hat a}_*}^{ - 2}{S^{ - 3/5}} \, .
\end{eqnarray}
A careful consideration of this equation reveals that the two terms on the right-hand-side must approximately balance each other. Therefore, we end up with
\begin{equation} \label{k_dominant}
{\hat k_ *} \simeq {c_k} {{\hat a}_*}^{ - 5/4}{S^{ - 1/4}} ,
\end{equation}
where ${c_k}$ is a $\mathcal{O}(1)$ coefficient. Then, using Eq. (\ref{k_dominant}), we obtain the expressions
\begin{equation} \label{gamma_delta_dominant}
{{\hat \gamma }_ *} \simeq {c_\gamma} {{\hat a}_*}^{-3/2} S^{-1/2} \, , \quad  {{\hat \delta }_{{\rm{in}}*}} \simeq {c_\delta} {{\hat a}_*}^{3/4} S^{-1/4} \, ,
\end{equation}
where ${c_\gamma}$ and ${c_\delta}$ are also $\mathcal{O}(1)$ coefficients. These relations show that the \emph{dominant} mode at the end of the linear phase exhibits the same scaling properties of the \emph{fastest growing} mode \cite{FKR_1963}, the latter of which satisfies the equation ${{\partial \hat \gamma /\partial \hat k} |_{{{\hat k}_{f*}},{{{\hat t}_*}}}} = 0$. This property arises because the additional contributions in Eq. (\ref{Fderdef}) can be shown to be negligible if ${{\hat w}_0} \ll {{\hat \delta }_{{\rm{in}}}} ({{\hat k}_*},{{\hat t}_0})$. If the current layer has the time to evolve until the Sweet-Parker inverse aspect ratio (${{\hat a}_*} \to S^{-1/2}$) is attained, then we obtain
\begin{equation} \label{SP_scalings}
{{\hat \gamma }_*} \simeq {c_\gamma} S^{1/4} \, , \quad {\hat k_*} \simeq {c_k} {S^{3/8}} \, , \quad {{\hat \delta }_{{\rm{in}}*}} \simeq {c_\delta} S^{-5/8} \, ,
\end{equation}
which exactly matches previous studies of the plasmoid instability \cite{TajimaShibata1997,Lou_2007,BHYR_2009,Baalrud_2012,Comisso2016} that were undertaken assuming a fixed Sweet-Parker current sheet.

However, for very high Lundquist numbers, the plasmoids complete their linear evolution well before the Sweet-Parker aspect ratio is reached. Therefore, we have to evaluate ${\hat a}_*$ for a more general case. This can be done by substituting the relations (\ref{k_dominant}) and (\ref{gamma_delta_dominant}) into Eq. (\ref{Eq_1a}), which give us the following {\it inverse-aspect-ratio equation}:
\begin{equation} \label{eq_3a} 
\ln \left( {{c_\delta }\frac{{\hat a_0^{1/2}}}{{{{\hat w}_0}}}\frac{{\hat a_*^{1/4}}}{{{S^{1/4}}}}} \right) = \frac{1}{2}\int_{{{\hat a}_0}}^{{{\hat a}_*}} {\hat \gamma (\hat a)\frac{{d\hat t}}{{d\hat a}}d\hat a} \, .
\end{equation}
This expression yields the final inverse aspect ratio ${\hat a}_*$ for a general current sheet evolution ${\hat a}({\hat t})$. It shows clearly that ${\hat a}_*$, and consequently the scaling relations of ${\hat t_*}$, ${{\hat \gamma }_*}$, ${\hat k_*}$ and ${{\hat \delta }_{{\rm{in}}*}}$, cannot be universal, as they must depend on the specific form of the function ${\hat a}({\hat t})$. 

We proceed further by considering, arguably, the most common case of current sheet thinning - the exponential thinning - which is typically the result of an instability-driven current sheet. Subsequently, we generalize the exponential thinning to encompass even algebraic cases. Other, more uncommon, possibilities could also be investigated, since the developed framework is fairly general.

We consider a current sheet width that shrinks in time as 
\begin{equation} \label{adef}
\hat{a}(\hat{t}) =  \hat{a}_0 e^{-\hat{t}/\tau} + \hat{a}_\infty \big( 1-e^{-\hat{t}/\tau} \big) \, ,
\end{equation}
where $\hat{a}_\infty = S^{-1/2}$. Note that the plasmoid width (\ref{half_width}) starts to grow only when $\hat \gamma > 1/\tau$. For ${\hat t_*} > \tau \ln (1 + {\hat a_0}{S^{1/2}})$ the current sheet approaches ${\hat a_*} \simeq {S^{-1/2}}$. Therefore, one can easily recover the relations (\ref{SP_scalings}). However, for the most interesting case ${\hat t_*} < \tau \ln (1 + {\hat a_0}{S^{1/2}})$ \cite{Note1}, which occurs for very large $S$-values, one has to solve Eq. (\ref{eq_3a}). 
In this case ${{\hat t}_*} \simeq \tau \ln ({{\hat a}_0}/{{\hat a}_*})$, and, using ${\hat a}_* \ll \hat{a}_0$, we obtain 
\begin{equation} \label{Implicit_a_final_exp}
{{\hat a}_*} \simeq {c_a} {\tau ^{2/3}}{S^{ - 1/3}}{\left[ {\ln \left( {{c_\delta }\frac{{\hat a_0^{1/2}}}{{{{\hat w}_0}}}\frac{{\hat a_*^{1/4}}}{{{S^{1/4}}}}} \right)} \right]^{ - 2/3}} \, ,
\end{equation}
where ${c_a} \approx 0.3$. This implicit equation for ${{\hat a}_*}$ can be solved by iteration using $\hat a_*^{(0)} \simeq {c_a}{\tau ^{2/3}}{S^{ - 1/3}}$ as the lowest order solution. Then, to the next order, we find
\begin{equation} \label{a_final_exp}
{{\hat a}_*} \simeq {c_a}\frac{{{\tau ^{2/3}}}}{{{S^{1/3}}}}{\left[ {\ln \left( {\frac{{{\tau ^{1/6}}}}{{{S^{1/3}}}}\frac{{\hat a_0^{1/2}}}{{{{\hat w}_0}}}} \right)} \right]^{ - 2/3}} \, .
\end{equation}
The logarithmic contribution to this expression is extremely important, since ${{\hat \gamma }_*}$ and ${\hat k_*}$ exhibit a super-linear dependence on $1/{\hat a}_*$, implying that the former two are strongly affected by changes in ${\hat a}_*$. Equation (\ref{a_final_exp}) reveals that ${\hat a}_*$ is \emph{smaller} than $\hat a_*^{(0)}$. Its value depends not only on $S$ but also on the perturbation amplitude ${\hat w}_{0} = 2 {\big( {{\hat \psi }_{0}}{{\hat a}_0} \big)^{1/2}}$ and the time scale of the driving process $\tau$.

\begin{figure}
\begin{center}
\includegraphics[bb = 2 30 358 231, width=8.6cm]{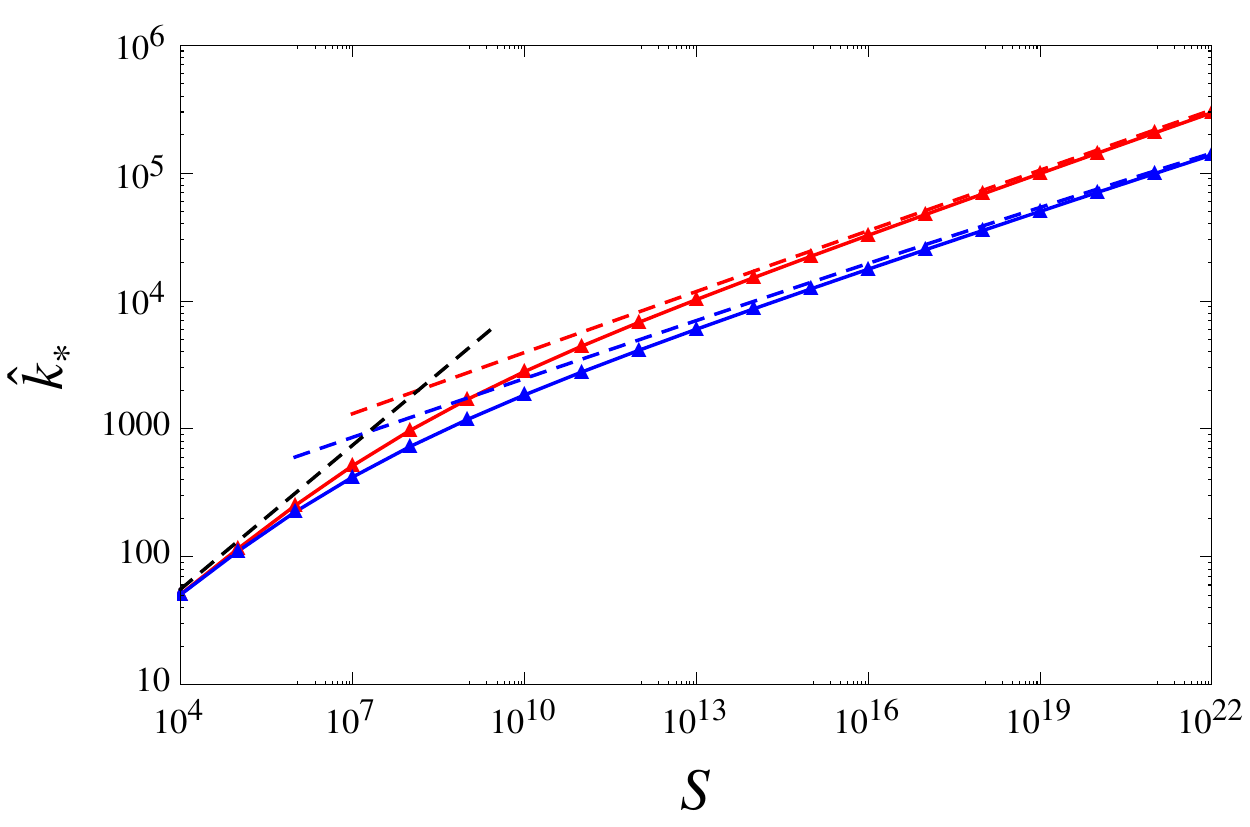}
\end{center}
\caption{${{\hat k}_*}$ vs. $S$ for ${\hat w}_0 = 10^{-12}$ (blue) and ${\hat w}_0 = 10^{-20}$ (red). In both cases $\tau = 1$ and ${\hat a}_0 = 1/2\pi$. Solid and dashed lines refer to numerical and analytical solutions, respectively. The black dashed line denotes ${{\hat k}_*} \sim S^{3/8}$.}
\label{fig1}
\end{figure}
\begin{figure}
\begin{center}
\includegraphics[bb = 2 30 358 231, width=8.6cm]{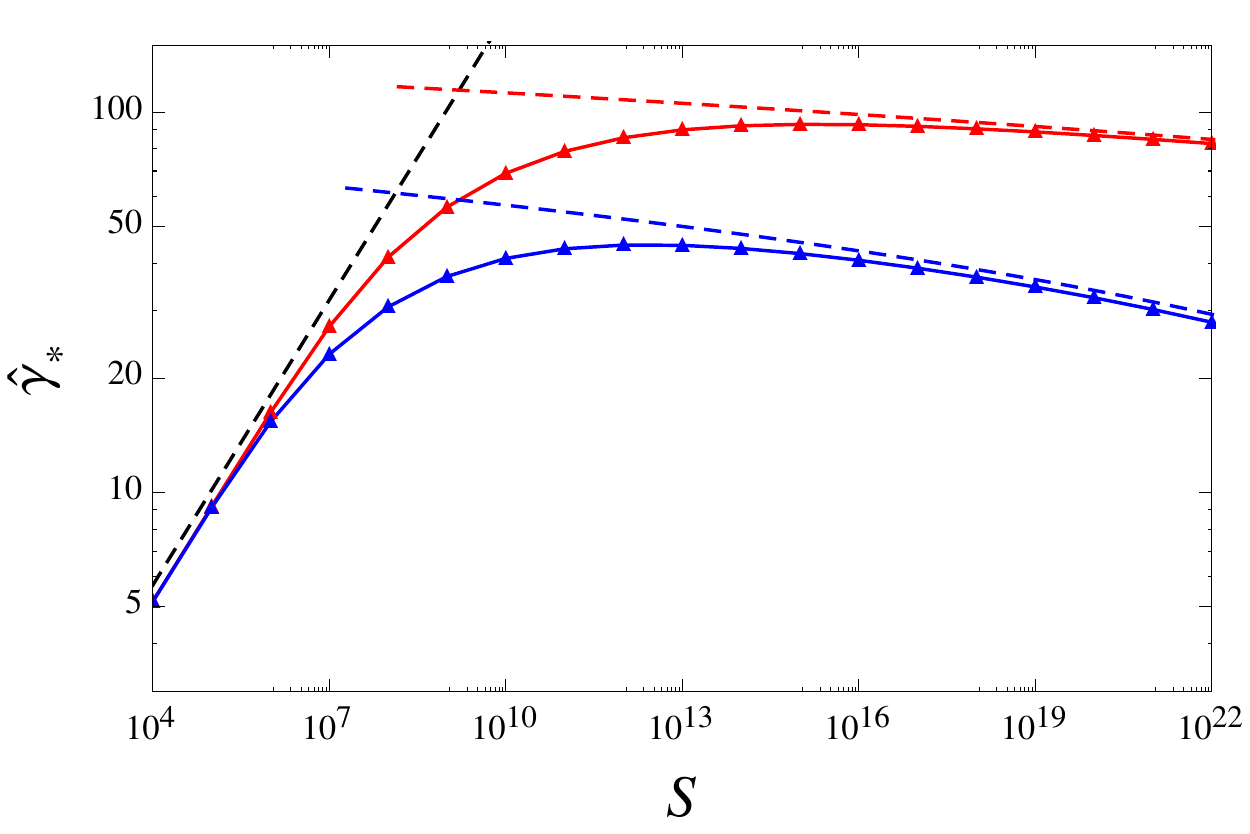}
\end{center}
\caption{${{\hat \gamma }_*}$ vs. $S$ for ${\hat w}_0 = 10^{-12}$ (blue) and ${\hat w}_0 = 10^{-20}$ (red). In both cases $\tau = 1$ and ${\hat a}_0 = 1/2\pi$. Solid and dashed lines refer to numerical and analytical solutions, respectively. The black dashed line denotes ${{\hat \gamma }_*} \sim S^{1/4}$.}
\label{fig2}
\end{figure}

Substituting Eq. (\ref{a_final_exp}) into Eqs. (\ref{k_dominant}) and (\ref{gamma_delta_dominant}), we duly obtain
\begin{equation} \label{k_final_exp}
{{\hat k}_*} \simeq {c_k}c_a^{-5/4} \frac{{{S^{1/6}}}}{{{\tau ^{5/6}}}}{\left[ {\ln \left( {\frac{{{\tau ^{1/6}}}}{{{S^{1/3}}}}\frac{{\hat a_0^{1/2}}}{{{{\hat w}_0}}}} \right)} \right]^{5/6}} \, ,
\end{equation}
\begin{equation} \label{gamma_final_exp}
{{\hat \gamma }_*} \simeq {c_\gamma }c_a^{ - 3/2}\frac{1}{\tau }\ln \left( {\frac{{{\tau ^{1/6}}}}{{{S^{1/3}}}}\frac{{\hat a_0^{1/2}}}{{{{\hat w}_0}}}} \right) \, ,
\end{equation}
\begin{equation} \label{delta_final_exp}
{{\hat \delta}_{{\rm{in}}*}} \simeq {c_\delta }c_a^{3/4}{\left( {\frac{\tau }{S}} \right)^{1/2}}{\left[ {\ln \left( {\frac{{{\tau ^{1/6}}}}{{{S^{1/3}}}}\frac{{\hat a_0^{1/2}}}{{{{\hat w}_0}}}} \right)} \right]^{ - 1/2}} \, .
\end{equation}
These relations are clearly very different from relations (\ref{SP_scalings}). It is instructive to compare our analytical solutions with the numerical solutions of the full principle of least time computed from Eqs. (\ref{Fermat_Principle}). The solutions for ${\hat k}_*$ and ${{\hat \gamma }_*}$ are shown in Figs.~\ref{fig1} and \ref{fig2} for two different values of ${{\hat w}_0}$. The analytical Sweet-Parker-limit solution is accurate only for moderately high $S$-values, while Eqs. (\ref{k_final_exp}) and (\ref{gamma_final_exp}) are increasingly exact for larger values of $S$. The behavior of ${{\hat \gamma }_*}$ is \emph{non-monotonic} in $S$, while ${\hat k}_*$ (proportional to the number of plasmoids) does exhibit a monotonic behavior, although it is lower when compared to the Sweet-Parker-based solution for large $S$.

It is important to consider the time scale of the plasmoid instability. 
From Eq. (\ref{a_final_exp}) it follows that
\begin{equation} \label{t_final}
{{\hat t}_*} \simeq \tau \ln \left\{ {\frac{{\hat a_0}}{{{c_a}}}\frac{{{S^{1/3}}}}{{{\tau ^{2/3}}}}{{\left[ {\ln \left( {\frac{{{\tau ^{1/6}}}}{{{S^{1/3}}}}\frac{{\hat a_0^{1/2}}}{{{{\hat w}_0}}}} \right)} \right]}^{2/3}}} \right\} \, .
\end{equation} 
Naturally, this time depends on the calibration of the ``clock'', i.e., on the starting time that is set by the initial inverse aspect ratio ${\hat a_0}$. While this constitutes a ready observable, it is not the \emph{intrinsic} time scale of the instability $\left({\tau_p}\right)$. This is because the mode ${{\hat k}_*}$ remains quiescent for an extended period of time, while it is subject to rapid growth only over a small fraction of ${{\hat t}_*}$.

The emergent mode effectively starts growing when its instantaneous growth rate is such that $\hat \gamma ({{\hat k}_*},{{\hat t}_{\rm{on}}}) > 1/\tau $. Therefore, upon using Eq. (\ref{fullgamma}) and retaining the dominant terms, we obtain ${{\hat a}_{\rm{on}}} \simeq {{\hat k}_*} {S^{ - 1/2}}{\tau ^{3/2}}$. Consequently, the intrinsic time scale of the plasmoid instability, defined via ${\tau _p} = \tau \ln ({{\hat a}_{\rm{on}}}/{{\hat a}_*})$, becomes
\begin{equation} \label{tauPfin}
{\tau_p} \simeq \tau \ln \left\{ {\frac{{{c_k}}}{{c_a^{9/4}}}{{\left[ {\ln \left( {\frac{{{\tau ^{1/6}}}}{{{S^{1/3}}}}\frac{{\hat a_0^{1/2}}}{{{{\hat w}_0}}}} \right)} \right]}^{3/2}}} \right\} \, .
\end{equation}
Because of the very weak dependence on $S$ (of the form $\ln \ln S$) and the perturbation amplitude, it is manifest that the intrinsic time scale of the plasmoid instability is near-universal for exponentially thinning current sheets.

\begin{figure}
\begin{center}
\includegraphics[bb = 2 30 358 231, width=8.6cm]{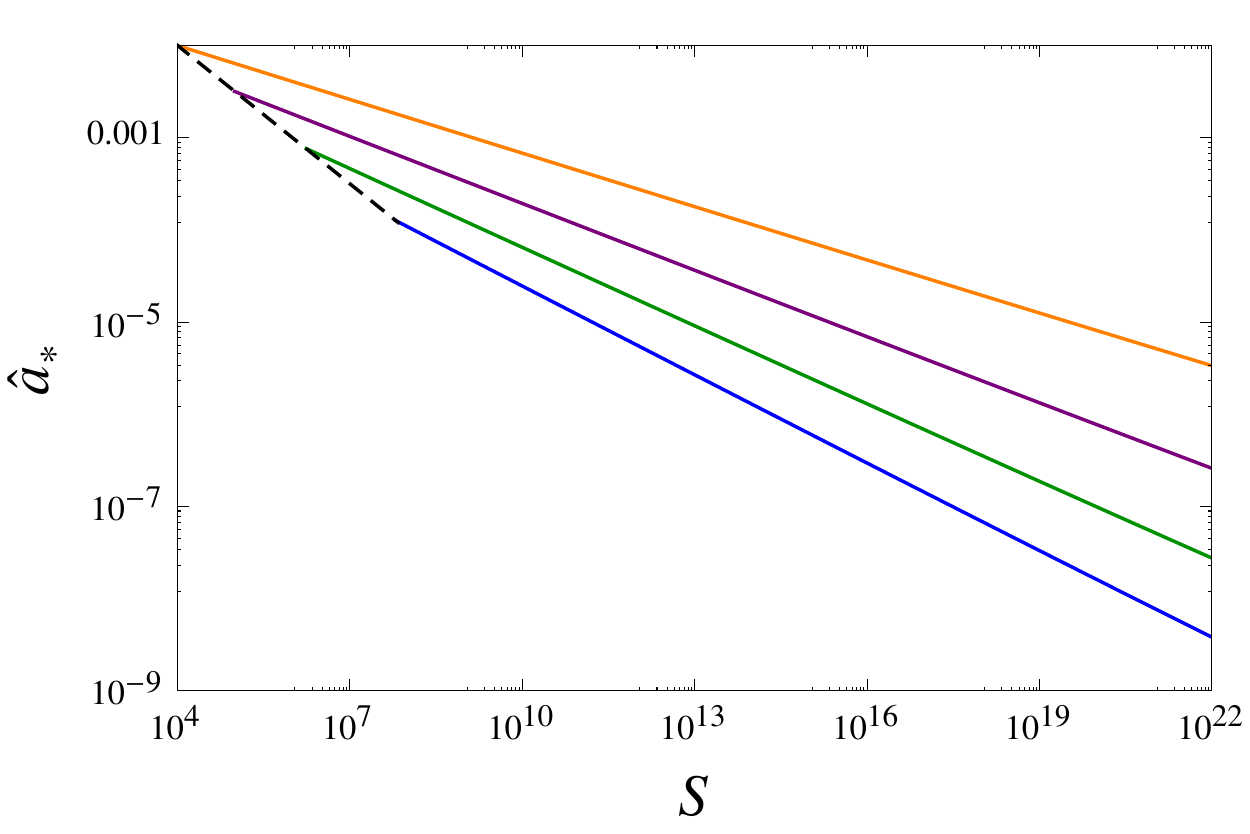}
\end{center}
\caption{${{\hat a}_*}$ vs. $S$ for $\tau = 1$, ${\hat w}_0 = 10^{-12}$, and ${\hat a}_0 = 1/2\pi$. Different colors refer to Eq. (\ref{afinalgeneral}) with $n=1$ (orange), $n=2$ (purple), $n=5$ (green), and $n \to \infty$ (blue). The Sweet-Parker scaling ${{\hat a}_*} \sim S^{-1/2}$ (black dashed) can be attained only for moderately high $S$-values.} 
\label{fig3}
\end{figure}

Hitherto, we have focused our attention on the exponential case on account of its commonality. However, other cases may occur, in which the current sheet thinning depends algebraically on time, as is possible for forced reconnection in a stable plasma \cite{Comisso2015,HK_1985,WB_1992,Fitz_2003,Ebrahimi2015}. Therefore, we consider a generalized current shrinking function of the form
\begin{equation} \label{adef3}
{\hat a}(\hat{t}) = \left( {{{\hat a}_0} - {{\hat a}_\infty }} \right){\left( {\frac{\tau }{{\tau  + {\hat t}/n}}} \right)^n} + {{\hat a}_\infty } \, ,
\end{equation}
which recovers the exponential thinning in the limit $n \to \infty$. Here, we proceed in a more straightforward way by exploiting the fact that ${{\hat \gamma }_*} \approx {{\hat \gamma }_f}({{\hat t}_*})$, where ${{\hat \gamma }_f}$ is the instantaneous growth rate of the fastest growing mode. This implies that we can approximate $\hat \gamma (\hat a)$ with ${{\hat \gamma }_s}(\hat a)$ in Eq. (\ref{eq_3a}). In the large ${\hat t}_*$ regime, we recover the Sweet-Parker-limit solution. Otherwise, for ${\hat t}_* < n \tau \big[(1 + {{\hat a}_0} S^{1/2})^{1/n} - 1 \big]$, we obtain a generalized version of Eq. (\ref{Implicit_a_final_exp}), using again ${{\hat \Delta}'} {\hat a} \simeq 2(\hat{k}{\hat a})^{-1}$ and ${\hat a}_* \ll \hat{a}_0$.
Evaluating the implicit equation along the same lines as Eq. (\ref{Implicit_a_final_exp}), we obtain
\begin{equation} \label{afinalgeneral}
{{\hat a}_*} \sim {\lambda^\nu} \hat a_0^{\nu /n}\frac{{{\tau ^\nu }}}{{{S^{\nu /2}}}}{\left[ {\ln \left( {\frac{{{\tau ^{\nu /4}}}}{{{S^{(2 + \nu )/8}}}}\frac{{\hat a_0^{(2n + \nu )/4n}}}{{{{\hat w}_0}}}} \right)} \right]^{ - \nu }} \, ,
\end{equation}
where $\nu := 2n/(2 + 3n)$ and $\lambda := c_k^{-2/5} n/(2 + 4n)$. For $n \to \infty$, we recover the same scaling as Eq. (\ref{a_final_exp}), while it is evident that different choices of $n$ lead to different expressions for ${{\hat a}_*}$. The $n$-dependence of ${{\hat a}_*}$ is illustrated in Fig. \ref{fig3}. We can see that ${{\hat a}_*}$ decreases for larger values of $n$, and that the domain of validity of the Sweet-Parker-limit solution reduces with decreasing $n$-values.

Equation (\ref{afinalgeneral}) enables us to obtain the scaling relations of the plasmoid instability for different cases of current sheet thinning. Let us focus on a current sheet thinning that is inversely proportional in time ($n=1$), which has been studied widely, especially in analyzing forced reconnection in Taylor's model \cite{Comisso2015,HK_1985,WB_1992,Fitz_2003,Comisso_2015JPP}. We find that 
\begin{equation} \label{afinaln1}
{{\hat a}_*} \sim  c_k^{ - 4/25}{\left( {\frac{{{{\hat a}_0}}}{6}} \right)^{2/5}}\frac{{{\tau ^{2/5}}}}{{{S^{1/5}}}}{\left[ {\ln \left( {\frac{{{\tau ^{1/10}}}}{{{S^{3/10}}}}\frac{{\hat a_0^{3/5}}}{{{{\hat w}_0}}}} \right)}  \right]^{ - 2/5}} \, .
\end{equation}
Therefore, substituting this relation into Eqs. (\ref{k_dominant}) and (\ref{gamma_delta_dominant}), we obtain ${{\hat k}_*} \sim {\cal L}_1^{1/2}/{({{\hat a}_0}\tau )^{1/2}}$, ${{\hat \gamma }_*} \sim {S^{-1/5}}{\cal L}_1^{3/5}/{({{\hat a}_0}\tau )^{3/5}}$, and ${{\hat w}_*} \sim {S^{ - 2/5}}{\cal L}_1^{ - 3/10}{({{\hat a}_0}\tau )^{3/10}}$, where $\mathcal{L}_1 := \ln \big({\tau ^{1/10}}\hat a_0^{3/5}/{S^{3/10}}{{\hat w}_0} \big)$. 
The final time is ${{\hat t}_*} \sim \tau {{\hat a}_0}/{{\hat a}_*}$, much higher than the exponential thinning case. These scaling laws are considerably different when compared to the latter - the plasmoid instability is less violent and a low number of plasmoids emerge in the nonlinear phase.

In this Letter 
we have generalized the previous treatments of the plasmoid instability by formulating a principle of least time for plasmoids in time evolving current sheets. We have shown that the scaling relations of all relevant parameters are dependent not only on $S$ but also on the perturbation amplitude and the characteristic rate of current sheet thinning. We also presented a detailed explanation of the dynamics of the plasmoid instability - the system remains quiescent for a certain period of time until the thinning reaches a critical value. Once this value has been attained, the plasmoid instability occurs on a short time scale, leading to explosive growth of the plasmoids. Thus, the developed theory sheds new light on the onset problem of fast magnetic reconnection, and also highlights the role of the current sheet thinning in determining the onset time. Direct numerical simulations supporting the theory will appear in a forthcoming paper.

A final remark concerning a universal feature of our results is in order. It is common in all realms of science to seek the existence of power laws, despite the fact that they are, sometimes, intrinsically simplistic \cite{Newman}. In contrast, we find that the scaling relations of the plasmoid instability are \emph{not} true power laws - a result that has never been derived or predicted before. \\

\begin{acknowledgments}
It is a pleasure to acknowledge fruitful discussions with Eero Hirvijoki, Hantao Ji, Russell Kulsrud, Roscoe White and Yao Zhou. 
This research was supported by the NSF Grant Nos. AGS-1338944 and AGS-1460169, and by the DOE Grant No. DE-AC02-09CH-11466.
\end{acknowledgments}

\end{document}